\documentclass[letterpaper, 10 pt, conference]{ieeeconf}
\IEEEoverridecommandlockouts

\usepackage{cite}
\usepackage{amsmath,amssymb,amsfonts}
\usepackage{pgfplots}
\usepackage{tikz}
\usepackage{tikzscale}
\usepackage{algorithmic}
\usepackage{balance}
\usepackage{graphicx}
\usepackage{textcomp}
\usepackage{xcolor}
\usepackage{subcaption}
\usepackage{optidef}
\usepackage[capitalise]{cleveref}
\usepackage{resizegather}
\crefname{equation}{}{} 
\Crefname{equation}{Equation}{Equations}

\def\BibTeX{{\rm B\kern-.05em{\sc i\kern-.025em b}\kern-.08em
    T\kern-.1667em\lower.7ex\hbox{E}\kern-.125emX}}

\pgfplotsset{compat=newest}
\usepgfplotslibrary{groupplots}
\usepgfplotslibrary{polar}
\usepgfplotslibrary{smithchart}
\usepgfplotslibrary{statistics}
\usepgfplotslibrary{dateplot}
\usepgfplotslibrary{ternary}
\usetikzlibrary{arrows.meta}
\usetikzlibrary{backgrounds}
\usepgfplotslibrary{patchplots}
\usepgfplotslibrary{fillbetween}
\usepackage{multirow}

\begin{document}

\title{\LARGE \bf Propulsion-Free Cross-Track Control of a LEO Small-Satellite Constellation with Differential Drag}

\author{Giusy Falcone$^{1}$, Jacob B. Willis$^{1}$, and Zachary Manchester$^{1}$
\thanks{$^{1}$Giusy Falcone, Jacob Willis, and Zachary Manchester are with the Robotics Institute,
        Carnegie Mellon University, 5000 Forbes Ave., Pittsburgh PA, 15213
        {\tt\small \{gfalcone, jwillis2, zmanches\}@andrew.cmu.edu}}%
}

\maketitle

\begin{abstract}
In this work, we achieve propellantless control of both cross-track and along-track separation of a satellite formation by manipulating atmospheric drag.
Increasing the differential drag of one satellite with respect to another directly introduces along-track separation, while cross-track separation can be achieved by taking advantage of higher-order terms in the Earth's gravitational field that are functions of altitude.
We present an algorithm for solving an n-satellite formation flying problem based on linear programming.
We demonstrate this algorithm in a receeding-horizon control scheme in the presence of disturbances and modeling errors in a high-fidelity closed-loop orbital dynamics simulation.
Our results show that separation distances of hundreds of kilometers can be achieved by a small-satellite formation in low-Earth orbit over a few months.
\end{abstract}


\section{Introduction}
Formations of multiple satellites are frequently used to perform tasks that a single satellite cannot accomplish alone. 
Examples include satellite navigation systems, like the global positioning system (GPS), and communications constellations like Iridium and Starlink.
The ability to maneuver and control the relative positions of such satellites is key to establishing and maintaining a formation.
Typically, such maneuvering requires a propulsion system and the consumption of propellant, the supply of which can be very limited, especially on smaller spacecraft.
In fact, many small spacecraft are not equipped with a propulsion system at all.

As an alternative to propulsion, external perturbation forces can be harnessed to affect the orbit of a satellite.
In low-Earth orbit (LEO), where most small satellites operate, two perturbation forces dominate satellites' orbital dynamics.

\begin{figure}[htb]
    \centering
    \includegraphics[width=0.8\columnwidth]{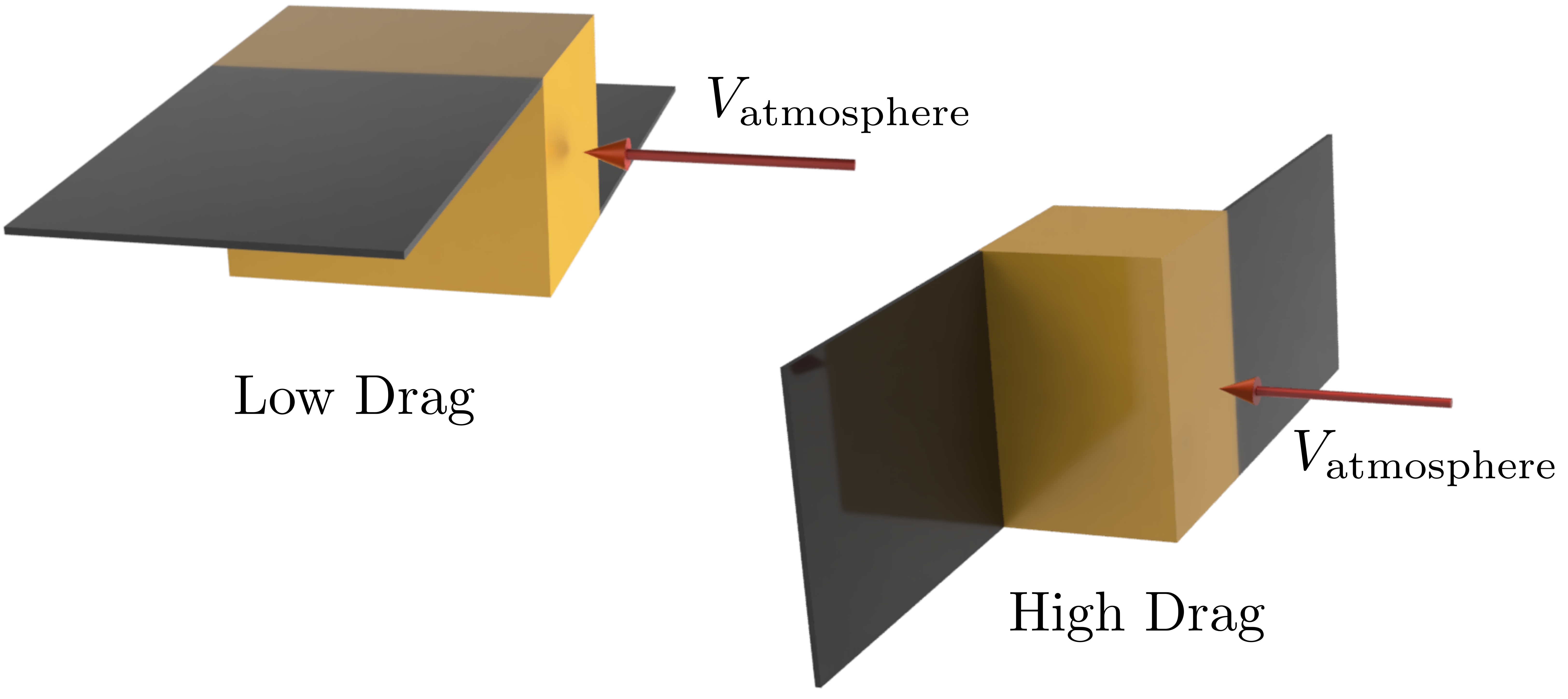}
    \caption{High and low drag configurations for a satellite with attitude-controlled drag modulation.}
    \label{fig:high_low_drag}
\end{figure}

The first perturbation force is atmospheric drag, which acts in the orbital plane and only directly influences the altitude of a satellite.
As drag changes a satellite's altitude, its orbital velocity and along-track position also change.
As depicted in \cref{fig:high_low_drag}, the drag area of a spacecraft can be changed by controlling the attitude of the spacecraft.
By placing some spacecraft in a high-drag state and others in a low-drag state, a differential drag between satellites can be introduced and the relative along-track positions of satellites can be changed.
On-orbit, this method has been used to establish and control the along-track positions for constellations of up to 100 satellites~\cite{foster_constellation_2018}.

The second perturbation force on a satellite in LEO is nodal precession.
Nodal precession is due to the Earth not being a perfect sphere and causes orbits to precess, or rotate, around the Earth's axis.
This effect introduces a small cross-track acceleration on a satellite that varies with altitude.
By establishing a large differential altitude between spacecraft, the nodal precession of those spacecraft will occur at different rates, and a cross-track orbital change can be made.

Most differential-drag formation flying methods ignore the cross-track influence of nodal precession because it is small compared to the along-track drift, requiring large altitude differences and long time horizons to have a significant effect.
In this paper, we describe a method for performing long-time-horizon differential drag maneuvers that utilize nodal precession to accomplish both along-track and cross-track changes in a formation configuration.
Our contributions include:
\begin{itemize}
    \item  An analytic expression for the first-order relationship between along-track and cross-track separation changes. This defines a fundamental limit on what along-track and cross-track separations are simultaneously achievable
    \item A convex trajectory optimization formulation to compute differential-drag sequences that achieve desired formation maneuvers
    \item A receding-horizon control strategy that re-plans maneuvers every few orbits to compensate for disturbances and modeling errors
    \item Simulation results demonstrating our receding-horizon controller performing several different maneuvers in a high-fidelity orbital dynamics simulation
\end{itemize}

The paper proceeds as follows:
In \cref{sec:related} we review previous research and on-orbit demonstrations of drag-based formation flying.
\Cref{sec:background} introduces background concepts that are used in the along-track and cross-track formation flying linear trajectory optimization that we develop in \cref{sec:formation}.
The results of a single convex trajectory optimization and closed loop simulations with the trajectory optimization as a feedback controller are shown and discussed in \cref{sec:experiments}.
We then conclude and comment on future work in \cref{sec:conclusion}.

\section{Related Work}\label{sec:related}
To avoid the need for a traditional propulsion system, many studies have investigated using drag modulation to control the relative positions of satellites in a formation. 

Leonard, et al. first proposed the use of drag modulation for maintaining the relative separation of spacecraft already in formation~\cite{leonard_orbital_1989}.
At a similar time, Mathews and Leszkiewicz~\cite{mathews_efficient_nodate} investigated a drag-propulsion combination to maintain a cyclical altitude and phase relationship between a spacecraft and a space station.
Additional methods for along-track formation keeping using drag have been proposed since then~\cite{kumar_differential_2011,hunter2022closed,varma_multiple_2012}.
Reconfiguration of a formation using differential drag or solar radiation pressure are considered by Spiller et al.~\cite{spiller_minimum-time_2017}.
They use the linear Hill-Clohessy-Wiltshire equations for the relative dynamics, so the formation size must remain small.

Differential drag control has also been studied in the context of along-track rendezvous.
Bevilacqua and Romano include $J_2$ perturbations in their model, but do not include cross-track separation in their relative state~\cite{bevilacqua_rendezvous_2008}.
They solve the drag-based rendezvous problem using a two-step analytic method.
An optimization approach to solving this problem was proposed by Harris and Açıkmeşe~\cite{harris_minimum_2014}.
They solve the problem as a constrained linear program with minimum-time cost.
Most differential drag methods assume binary drag states where a satellite is in either a low or high drag configuration.
Harris et al. investigate a continuous drag modulation scheme based on the coupling of spacecraft attitude and drag~\cite{harris2020linear}.

There have been multiple demonstrations of differential-drag control on orbit.
The ORBCOMM communications constellation, launched in 1997-1999, used modulation of differential drag, along with occasional propulsive maneuvers, to maintain the along-track separation for their network of thirty spacecraft~\cite{maclay2005satellite}.
Each week a new plan for the ORBCOMM constellation was created that oriented the solar arrays in high or low drag configurations during the eclipse phase of each orbit.
On orbit, the system performed within mission parameters for ten years.
A limited demonstration of differential drag modulation using deployable panels was performed on-orbit by the AeroCube-4 CubeSat mission in 2012~\cite{gangestad2013operations}.

Perhaps the most well-known and complete on-orbit demonstration of differential drag was for the Planet Earth-imaging constellation~\cite{foster_constellation_2018}.
The Planet constellation performed both the initial slot allocation and phasing of satellites as well as station keeping of satellites using differential drag.
After deployment and initial contact, the slot allocation and phasing problem was solved by a ground control system using a genetic algorithm.
This used the initial differences in satellite position to reduce the phasing time.
In its original form, the dynamics are written as a two-dimensional linear system using the Gauss variational equations and the drag-control is considered binary.
A continuous optimization of the Planet slot allocation and phasing problem was formulated by Blatner~\cite{blatner_optimal_2018}.
Repeated updates to handle perturbations, and continuous controls were presented by Sin et al.~\cite{sin_small_2018}.

The CYGNSS constellation also included differential-drag modulation for along-track phasing in its mission design~\cite{bussy-virat_assessment_2019}.
The control design includes operational constraints for sun pointing of solar panels and nadir pointing of the science instruments.
The constellation was deployed at a higher than expected altitude, resulting in lower drag than was anticipated, and the differential drag maneuvers were ineffective within the initial mission timeframe.
An adaptive Lyapunov controller for handling uncertain drag conditions on-line is presented in~\cite{perez_differential_2016}.

None of the previously discussed works consider large out-of-plane or cross-track motion of the satellites.

Using nodal precession to modify the cross-track separation of a formation was demonstrated on-orbit by the FORMOSAT-3/COSMIC mission~\cite{fong_constellation_2008}.
For this mission, propulsion was used to raise the orbital altitude of some satellites, the cross-track separation was allowed to increase for a period of time, and the orbital altitudes of all satellites were then matched to eliminate drift.
Similar studies have extended these propulsive methods to non-circular orbits~\cite{mahdisoozani_developing_2021}, low-thrust~\cite{mcgrath_general_2020}, and handling the perturbing affects of drag~\cite{di_pasquale_optimization_2022}.

Two works~\cite{leppinen_deploying_2016,lee2022ground} combine differential drag and nodal precession to modify the cross-track separation of satellites.
These works are the most similar to ours.
Leppinen~\cite{leppinen_deploying_2016} performs a feasibility study and demonstrates that it is possible to obtain a sufficient altitude separation for nodal precession to change the RAAN of a satellite. 
No control methods are presented.
Lee and Bang~\cite{lee2022ground} present a method for modifying the ground-tracks of satellites in a constellation using differential drag and nodal precession.
They use a series of processes to solve both the slot allocation and phasing problems.
Synchronization of the along-track and cross-track state of the satellites is not investigated in either of these prior works.
This is a key contribution of our work.

There has been interest in utilizing aerodynamic lift for modifying a satellite's cross-track trajectories~\cite{horsley_small_2013,ivanov_study_2018}.
However, the lift force is at maximum an order of magnitude smaller than drag for a typical spacecraft, and for a symmetric spacecraft the lift often averages to zero~\cite{horsley_small_2013}, so we do not consider its effect here.

We formulate the differential-drag control problem as a convex trajectory optimization with a linear cost and linear constraints.
Tillerson, et al. solved spacecraft formation flying problems with convex trajectory optimization over twenty years ago~\cite{tillerson2002co}.
Since that time, convex trajectory optimization has gained popularity for solving many aerospace problems including orbital maneuvering, rocket soft landing, and planetary aerocapture~\cite{liu2017survey,malyuta2021advances}.
In all of these domains, convex trajectory optimization provides the advantage of solving highly constrained control problems in a computationally tractable manner.

\section{Background}\label{sec:background}
\subsection{Keplerian Motion}
The unperturbed Keplerian dynamics of a satellite orbiting around the Earth are described by the two-body equation
\begin{equation}
    \ddot{\mathbf{r}} = -\dfrac{\mu}{r^3} \mathbf{r}
    \label{eq:kep_eq}
\end{equation}
where $\mathbf{r}$ is the position vector of the spacecraft in the Earth-centered inertial frame, $\ddot{\mathbf{r}}$ is the acceleration vector and $\mu$ is the Earth's gravitational parameter.

\Cref{eq:kep_eq} assumes that the spacecraft is influenced only by the spherically symmetric gravitational field of the Earth. 
In reality, a spacecraft experiences a large number of secondary perturbation forces.
The largest perturbation forces on a satellite in LEO are due to atmospheric drag and the Earth's non-spherical gravitational field.  
To account for perturbations, \cref{eq:kep_eq} can be written as
\begin{equation}
    \ddot{\mathbf{r}} = -\dfrac{\mu}{r^3} \mathbf{r} + \mathbf{p}
    \label{eq:kep_eq_p}
\end{equation}
where $\mathbf{p}$ is the perturbative acceleration vector of the satellite. 

\begin{figure}[htb]
    \centering
    \includegraphics[width=0.7\columnwidth]{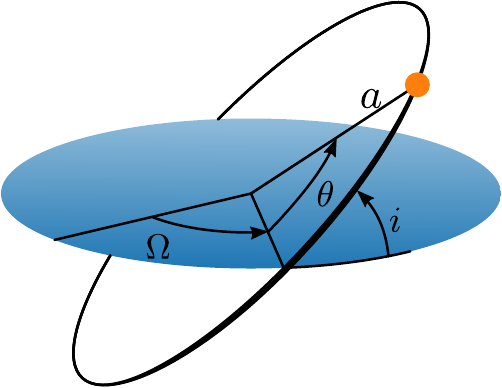}
    \caption{Notation used to describe the orbital state of a satellite in a circular orbit. Here, the blue plane is the Earth's equatorial plane and $\Omega$ is referenced to an inertially fixed direction.}
    \label{fig:notation}
\end{figure}

The orbital state of a satellite is commonly described using six quantities known as the orbital elements~\cite{curtis_2005}. The orbital elements are: $a$, the semi-major axis; $e$, the eccentricity of the orbit ellipse; $i$, the inclination; $\Omega$, the right ascension of the ascending node (RAAN); $\omega$, the argument of periapsis; and $\nu$, the true anomaly.
In this work we consider circular orbits, so $e=0$ and $\omega$ is undefined; since $\nu$ is referenced from $\omega$, it is poorly defined.
Instead, we use $\theta$, the argument of latitude (AoL), which measures along-track orbital position from the equatorial plane.
In the remainder of this work our focus will be on the dynamics of $a$, $\Omega$, and $\theta$.
The dynamics of the other orbital elements are minor with respect to drag and nodal precession.
\Cref{fig:notation} shows how $a$, $i$, $\Omega$, and $\theta$ describe the orbital state of a satellite in a circular orbit.

\subsection{Atmospheric Drag}

In LEO, atmospheric drag is modeled by
\begin{equation}
    \mathbf{D} = -\dfrac{1}{2 m}\rho A C_D v (\mathbf{v}-\mathbf{v_{atm}})\\
    \label{eq:drag}
\end{equation}
where $\mathbf{D}$ is the drag force, $\rho$ is the atmospheric density, $A$ is the satellite's incident cross-sectional area,  $C_D$ is the drag coefficient, $m$ is the satellite mass, $\mathbf{v}$ is the inertial velocity vector of the satellite, $\mathbf{v_{atm}}$ is the velocity of the atmosphere, and $v = \|\mathbf{v-v_{atm}}\|$ is the relative velocity vector magnitude~\cite{prussing_conway_2013, chobotov_2002}.

According to \cref{eq:drag}, the drag force increases with atmospheric density; however, it can also be modulated by changing the cross-sectional area of the satellite. 
The cross-sectional area can be modified either through deployable panels or by changing the spacecraft attitude~\cite{falcone2021energy,foster_constellation_2018}.

Drag always acts in the direction opposing velocity, and therefore can only directly affect the motion of a spacecraft within the orbital plane, decreasing its total orbital energy.
In the orbital plane, drag enters the dynamic equations for eccentricity and the semi-major axis.
Drag circularizes an orbit, decreasing its eccentricity~\cite{wertz_2005}; since we are assuming circular orbits, we do not consider the eccentricity dynamics due to drag here.
The semi-major axis dynamics due to drag are
 \begin{equation}
    \dot{a} = 2\sqrt{\dfrac{a^3}{\mu}}D 
\end{equation}
where $D = \|\mathbf{D}\|$ is the magnitude of the drag vector.  

\subsection{Nodal Precession and The Method of Averaging}

As discussed previously, the Earth is not a perfect sphere, but resembles an oblate spheroid; this causes the gravitational field of the Earth to deviate from the point mass model in \cref{eq:kep_eq}.
Models of the Earth's gravitational field are expressed by a spherical harmonic expansion with coefficients $J_*$~\cite{curtis_2005,chobotov_2002}.
We use a simple model based on the first non-trival term, $J_2$.
It captures the dominating effect of the Earth's oblateness and is three orders of magnitude larger than the next spherical harmonic term.
In addition, the $J_2$ model is rotationally symmetric around the Earth so it only requires knowledge of an orbit's inclination.
This makes it consistend across all possible RAAN angles and separations in a formation.

On short timescales, the $J_2$ perturbation affects all the orbital elements. 
However, on long timescales it introduces only a mean variation on $\Omega$, and averages to zero for $\theta$.
We denote the mean RAAN $\bar{\Omega}$ and the mean AoL $\bar{\theta}$. 
Their mean dynamics are
\begin{align}
    \dot{\bar{\Omega}} &= - \left [\dfrac{3}{2} \dfrac{J_2 \sqrt{\mu} R_E^2}{a^{7/2}} \right ] \cos i 
    \label{eq:omega}\\
    \dot{\bar{\theta}} &= \sqrt{\mu/a^3}
    \label{eq:theta}
\end{align}
where $R_E$ is the Earth's equatorial radius.

\section{Formation Flying}\label{sec:formation}
From \cref{eq:omega,eq:theta}, the nodal precession rate and the AoL rate are both functions of the semi-major axis.
The semi-major axis can be modulated through drag variation~\cref{eq:drag}.
This means that drag modulation can be used to affect changes in the AoL and RAAN for a satellite and establish satellite formations with coplanar and non-coplanar separations.
In this section we develop the linear dynamics model and the trajectory optimization method we use for drag-based formation flying.

\subsection{Linearized Dynamics}
Using a first-order Taylor expansion, \cref{eq:omega,eq:theta} can be linearized with respect to a reference semi-major axis, $a$, and \cref{eq:drag} can be linearized with respect to a reference drag, $D$, as follows
\begin{align}
\label{eq:theta_lin}
    \Delta\dot{\bar{\theta}} &= -\dfrac{3}{2}\sqrt{\dfrac{\mu}{a^5}}\Delta{a} \triangleq k_1 \Delta{a}\\ 
    \label{eq:a_lin}
    \Delta\dot{a} &= 2\sqrt{\dfrac{a^3}{\mu}}D \Delta{D} \triangleq k_3 \Delta{D}\\ 
    \label{eq:omega_lin}
    \Delta\dot{\bar{\Omega}} &= \dfrac{21}{4}J_2\sqrt{\dfrac{\mu}{a^9}}R^2_E \cos{i} \Delta a \triangleq k_2 \Delta a.
\end{align}

According to \cref{eq:theta_lin,eq:omega_lin}, the rates of $\Delta \bar{\theta}$ and $\Delta \bar{\Omega}$ are both governed by $\Delta a$, so they cannot be altered independently of each other.
Assuming $\Delta \bar{\theta} = \Delta \bar{\Omega} = 0$ initially, any $\Delta \bar{\theta}$ achieved results in
\begin{equation}
\label{eq:sep}
\Delta \bar{\Omega} = \dfrac{k_2}{k_1} \Delta \bar{\theta} \triangleq k_4 \Delta \bar{\theta} 
\end{equation}
where $k_4 = k_2 / k_1$ is a dimensionless constant that depends only on the reference orbit.

The linear equations in \cref{eq:theta_lin,eq:a_lin,eq:omega_lin} can be put in the standard form of a linear dynamical system,
\begin{equation}
\mathbf{\dot{x}} = A\mathbf{x}+B\mathbf{u}\\
\label{eq:lineq}
\end{equation}
where
\begin{equation} \label{eq:lin_xAB}
\mathbf{x} = \begin{bmatrix}
\Delta \bar{\theta}\\ 
\Delta a\\ 
\end{bmatrix}, A = \begin{bmatrix}
 0 & k_1 \\ 
 0 & 0  \\ 
\end{bmatrix},  B = \begin{bmatrix}
0\\ 
k_3 \\ 
\end{bmatrix}
\end{equation}
and $\mathbf{u} = \begin{bmatrix}
\Delta{D}
\end{bmatrix}.$
We omit $\Delta \bar{\Omega}$ from the state since \cref{eq:sep} establishes a relationship between $\Delta \bar{\Omega}$ and $\Delta \bar{\theta}$.

To extend this method to the case of $n>2$ satellites the first satellite is chosen as the chief satellite, and the other satellite's $\Delta$ states are all referenced to the chief.
We concatenate $n-1$ copies of \cref{eq:lin_xAB} to rewrite \eqref{eq:lineq} as a $2(n-1)$ state system.
When referring to the relative state between the chief and another satellite, we use the notation $\Delta a^{1-p}$ and $\Delta \bar{\theta}^{1-p}$, where $p$ is the index of the satellite.



\subsection{Constraints on the Final Conditions of Drag-Based Formation Control}
Given a pair of satellites deployed at the same initial orbit (i.e. $\mathbf{x}_0=0$), our goal is to manipulate the differential drag $\Delta D$ over time to achieve a final formation configuration $\mathbf{x}_f$ at some future time $t_f$. 
The basic control strategy is to lower the orbital altitude of one satellite such that its nodal precession rate is larger than the other satellite. 
The satellites then remain in this configuration, with $\Delta D = 0$ until a desired $\Delta \bar{\theta}$, and therefore a desired $\Delta \bar{\Omega}$, is achieved, at which time the higher satellite lowers its altitude to match the first satellite.
To maintain a fixed final formation configuration, we must have $\dot{\mathbf{x}}_f = 0$. 
To satisfy this, \cref{eq:theta_lin,eq:a_lin,eq:omega_lin} shows that $\Delta a_f$ and $\Delta D_f$ must be zero --- the satellites must be at the same final altitude and in the same drag configuration. 

Modifying \eqref{eq:sep} to account for the fact that $\Delta \bar{\theta}$ is an angular quantity, the possible $\Delta \bar{\Omega}$ for a desired final $\Delta\bar{\theta}_f$ are given by
\begin{equation}
    \label{eq:sep_f}
    \Delta\bar{\Omega}_f = k_4 (\Delta \bar{\theta}_f + 2 \pi \ell)
\end{equation}
where $\ell$ is any integer. 
To first order, \cref{eq:sep_f} defines the argument of latitude and right ascension separations achievable using drag modulation.
For differential-drag formation control to be feasible, \cref{eq:sep_f} is a fundamental limit that must be obeyed when selecting the final $\Delta \bar{\theta}$ and $\Delta \bar{\Omega}$ of a formation.

\subsection{Optimization-Based Drag Maneuver Planning}
Given  $n$ satellites deployed in the same orbit (i.e., $\mathbf{x}_0=0$), we desire to maneuver these satellites into a formation configuration at a final time $t_f$.
To do so with differential drag, we must choose the final state $\mathbf{x}_f$ by choosing the desired value for either $\Delta \Omega_f$ or $\Delta \theta_f$ and selecting the other in accordance to \cref{eq:sep_f}.
The final altitude or final time are then a result of this choice.
It then remains to find the necessary control inputs to achieve this formation.

A full trajectory of drag modulation inputs that drives the satellite formation from $\mathbf{x}_0$ to $\mathbf{x}_f$ can be planned by solving the convex optimization problem	
\begin{mini}<b> 
  {\mathbf{x}_{1:N}, \mathbf{u}_{1:N-1}}
  {g_f(\mathbf{x}_N) + \sum_{i=1}^{N-1} g(\mathbf{x}_i, \mathbf{u}_i)}
  {}{}
  \addConstraint{}{\mathbf{x}_{i+1} = A \mathbf{x}_i + B \mathbf{u}_i}{} \labelOP{opt:trajopt}
  \addConstraint{}{\left[\Delta a^{1-2}_N, ..., \Delta a^{1-n}_N   \right ] =0 }{}
  \addConstraint{}{\Delta a_\mathrm{min} \leq \left [\Delta a^{1-2}_i, ..., \Delta a^{1-n}_i   \right ] \leq \Delta a_\mathrm{max}}{}
  \addConstraint{}{u_\mathrm{min} \leq \mathbf{u}_i \leq u_\mathrm{max}}{}
\end{mini}
where $g(x,u)$ is a convex stage cost function, and $g_f(x)$ is a convex terminal cost function.
The first constraint enforces the discretized form of the linear dynamics from \cref{eq:lineq},
the second constraint ensures the satellites end at the same final altitude, 
the third constraint restricts the minimum and maximum altitude differences for each couple of satellites to be within $\Delta a_\mathrm{min}$ and $\Delta a_\mathrm{max}$,
and the final constraint enforces $u_\mathrm{min}$ and $u_\mathrm{max}$ as lower and upper bounds on the drag achievable by each satellite.
In this work, meeting the $\Delta\theta$ final conditions is not treated as a constraint but included in the cost function; this relaxes the problem and avoids ill-conditioning. 

The cost functions $g$ and $g_f$ can be chosen to shape the overall system behavior. To produce minimum-time bang-bang control commands, $L_1$ costs can be used~\cite{Zadeh196245}:

\begin{equation}\label{eq:costfunc}
\begin{gathered}
g_f(\mathbf{x}_N) = \left \| \Delta\bar{\theta}_N^{1-2} - \Delta\bar{\theta}_f^{1-2} \right \|_1 + ... + \left \| \Delta\bar{\theta}_N^{1-n} - \Delta\bar{\theta}_f^{1-n} \right \|_1 \\ 
g(\mathbf{x}, \mathbf{u}) = \left \| \Delta\bar{\theta}^{1-2} - \Delta\bar{\theta}_f^{1-2} \right \|_1 + ... + \left \| \Delta\bar{\theta}^{1-n} - \Delta\bar{\theta}_f^{1-n} \right \|_1 \\ + \left \| u_1 \right \|_1 + \left \| u_2 \right \|_1 +  ...+ \left \| u_n \right \|_1.
\end{gathered}
\end{equation}
Other convex cost functions, such as a quadratic cost, are also possible. 


\section{Simulation Experiments}\label{sec:experiments}

The convex optimization problem in \cref{opt:trajopt} with the cost function in \cref{eq:costfunc} is a linear program and can be solved with many standard solvers such as ECOS~\cite{ECOS}, GLPK~\cite{GLPK}, or MOSEK~\cite{MOSEK}. 
In these experiments, \cref{opt:trajopt} and \cref{eq:costfunc} are implemented in Julia using the Convex.jl modeling toolbox~\cite{convexjl} and solved with the MOSEK solver.

In all of our experiments we consider a constellation in which each satellite is a $1.5 \mathrm{kg}$ CubeSat with a $15 \mathrm{cm} \times 10 \mathrm{cm} \times 10 \mathrm{cm}$ chassis and equipped with two deployable solar panels each with dimension $20 \mathrm{cm} \times 15 \mathrm{cm} \times 0.3 \mathrm{cm}$.
A notional model of the satellite and its high and low drag configurations is shown in \Cref{fig:high_low_drag}.
The drag ratio of the satellite is 7.5:1.
To be conservative we use a 5:1 drag ratio here, which results in setting the input limits, $u_\mathrm{min}$ and $u_\mathrm{max}$, of \cref{opt:trajopt} to 0.2 and 1, respectively. 

\subsection{Trajectory Optimization}\label{sec:lp_results}

Here we solve \cref{opt:trajopt} for a pair of satellites deployed at 440km altitude and with an inclination of $51.5^\circ$ --- conditions that approximate deployment from the International Space Station (ISS).
The final conditions are set to $\Delta \bar{\theta}_f = 0$ and $\ell = 2$. 
From \cref{eq:sep_f}, this results in $\Delta \bar{\Omega}_f = 1.4^\circ$, a spherical distance of 165 km. 
In this scenario, the altitude limits, $\Delta a_\mathrm{max}$ and $\Delta a_\mathrm{min}$, are set to $\pm 10 \mathrm{km}$. 

\begin{figure}[htb]
    \centering
    \includegraphics[width=\columnwidth]{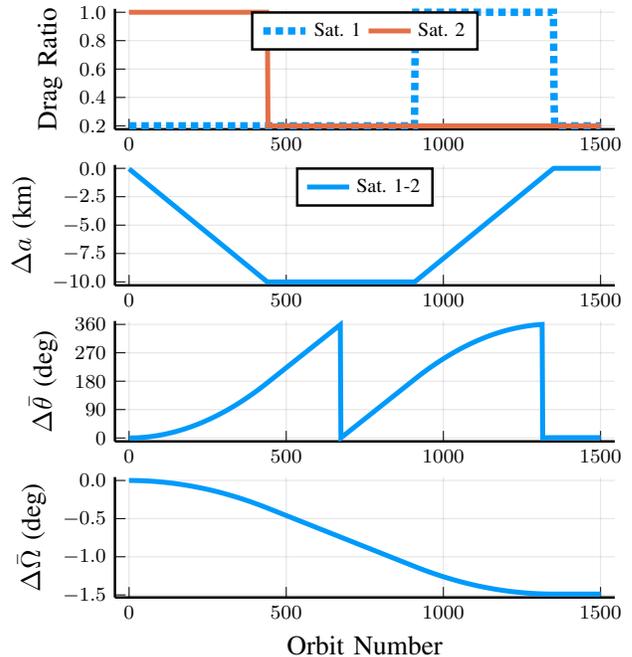}
    \caption{Linear trajectory optimization solution for a two-satellite formation. Top: the drag ratios for the satellites. Second: the relative altitude between the satellites. Third: the relative argument of latitude between the two satellites. Bottom: the relative right angle of the ascending node between the two satellites. The satellites end at the same altitude, resulting in a constant final argument of latitude and right angle of the ascending node.}
    \label{fig:lp_combined}
\end{figure}

The results of a single, 1500 orbit time horizon, trajectory optimization for these conditions are shown in \cref{fig:lp_combined}.
The top plot shows the drag control trajectory for the two satellites.
To increase the relative AoL and RAAN, the orbital altitude of the second satellite is decreased first. 
The bottom three plots show the change in $\Delta a$, $\Delta \bar{\theta}$, and $\Delta \bar{\Omega}$ respectively.
From these plots we can see that the the relative AoL increases by $720^\circ$, or two full orbits, and that the first satellite lowers its altitude to exactly reach $\Delta \bar{\theta}_f = 0$.
We can also see that the $10 \mathrm{km}$ altitude constraint was satisfied.
This optimization took $0.8 \mathrm{s}$ to solve on a MacBook Pro with an Apple M1 Pro processor.

\subsection{Closed-Loop Simulation Results}\label{sec:simulation}
To capture the effect of realistic  modeling errors and disturbances, closed-loop simulations are performed using nonlinear dynamics with additional perturbations not modeled in \eqref{eq:lineq}. 
Specifically, we incorporate the effects of Earth's rotation on drag, the first five zonal harmonics ($J_1$-$J_6$) for gravity, and a small eccentricity, $e = 0.005$, of the initial orbit. 
To successfully execute planned maneuvers in the presence of modeling errors and disturbances we compute solutions to \cref{opt:trajopt} in a receding-horizon or model-predictive control (MPC) loop. 
On each loop, the optimization problem is solved using the current measured state of the spacecraft formation, and the computed control inputs are applied over the next timestep.
The repeated solving corrects for disturbances on the spacecraft.
The timesteps of the control loop are on the order of one to five orbital periods.

\begin{figure}[htb]
    \centering
\begin{subfigure}[t]{0.95\columnwidth}
    \centering
    \includegraphics[width=\columnwidth]{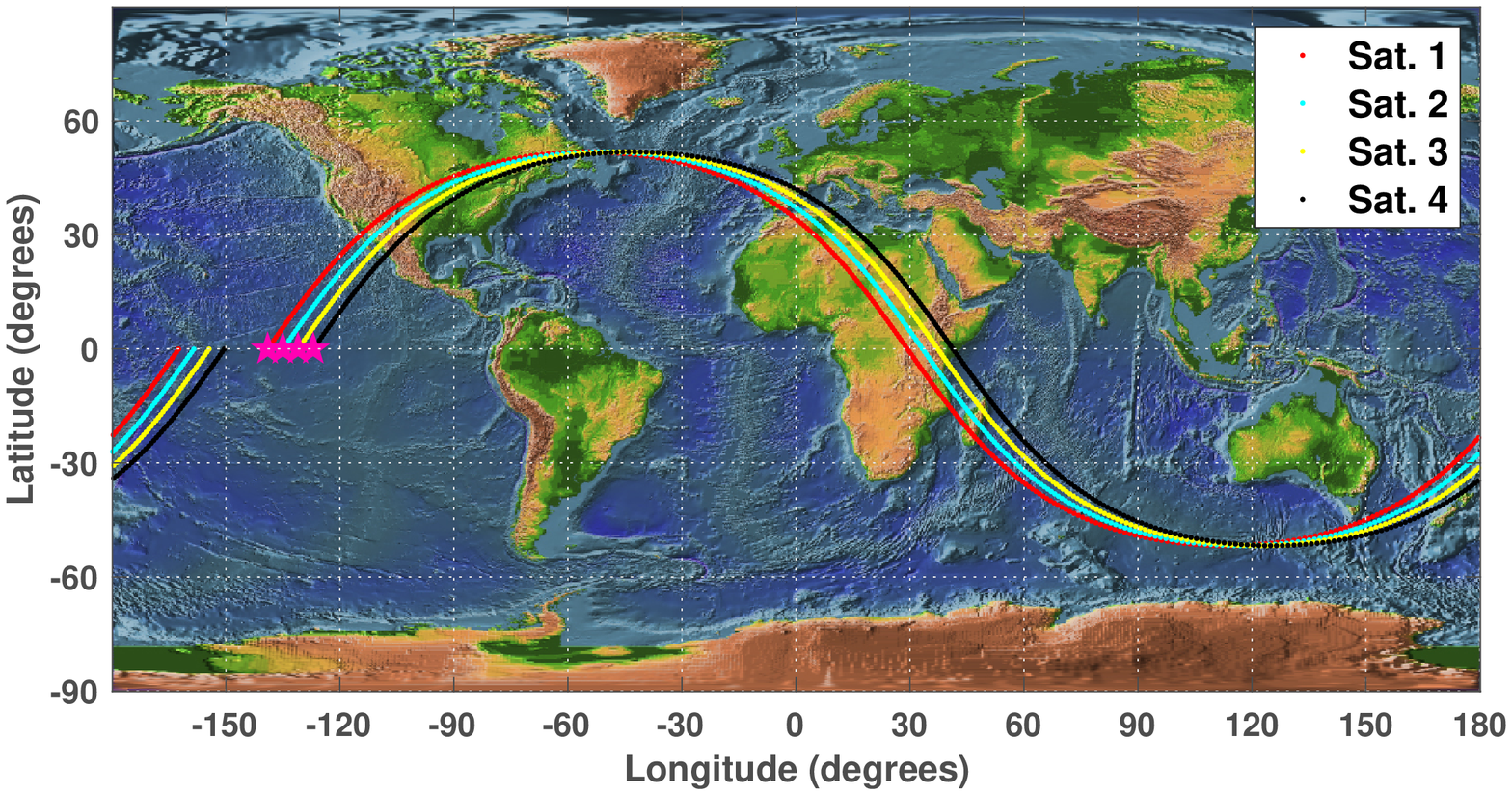}
    \caption{}
    \label{fig:scenario1}
    \end{subfigure}
\begin{subfigure}[t]{0.95\columnwidth}
    \centering
    \includegraphics[width=\columnwidth]{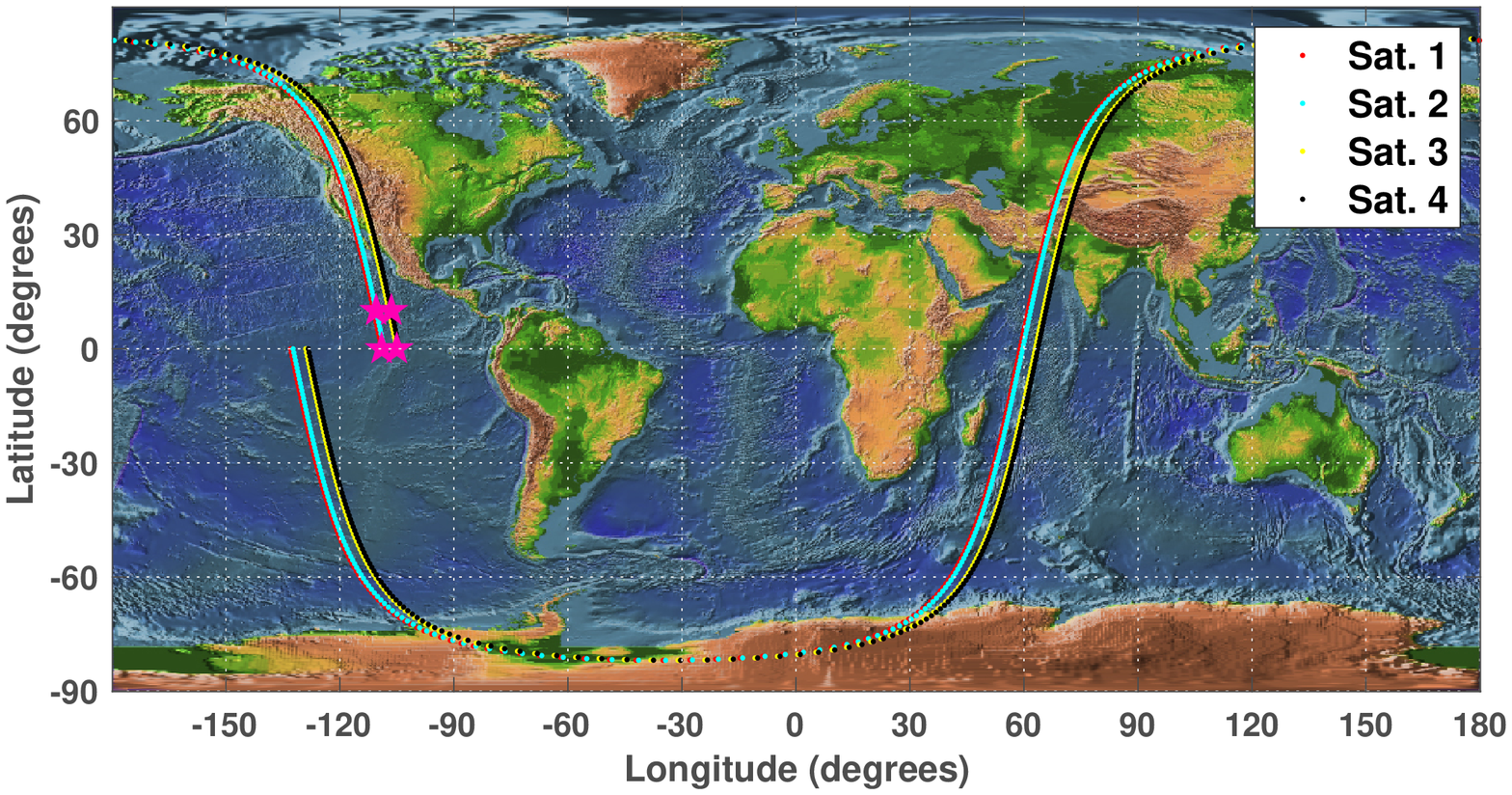}
    \caption{}
    \label{fig:scenario2}
        \end{subfigure}
    \caption{(a) Scenario 1: formation of four satellites in a line with equally distributed right angle of the ascending node. (b) Scenario 2: formation of four satellites distributed in argument of latitude and right angle of the ascending node to form the vertices of a square. }
\end{figure}

This receding-horizon control algorithm has been used to solve two scenarios, depicted in \cref{fig:scenario1,fig:scenario2} and described in \cref{sec:mpc_line,sec:mpc_square}.
The initial state is chosen to approximate deployment with two common CubeSat deployment methods: the ISS and a SpaceX Transporter launch.
Both scenarios assume that all of the satellites have the same drag ratio and the same initial state.

\subsubsection{Scenario 1 --- Line Formation}\label{sec:mpc_line}
Scenario 1 assumes that four satellites are deployed from the ISS, with an altitude of $440 \mathrm{km}$, eccentricity of $0.005$, and inclination of $51.5^\circ$.
The goal is to maneuver the satellites to be equally distributed in the cross-track direction with zero change in AoL so they pass over the equator in a line, as depicted in \cref{fig:scenario1}.
This corresponds to $\Delta \bar{\theta}_f = 0$ and $\Delta \bar{\Omega}_f = k_4 2 \pi \ell$ with $\ell = 1, 2, 3$.
In this scenario, the receding-horizon control policy is re-solved once per orbit over a time horizon of 1400 orbits and the altitude limits $\Delta a_{max}$ and $\Delta a_{min}$ were set to $\pm$100km.

\begin{figure}[htb]
    \centering
    \includegraphics[width=\columnwidth]{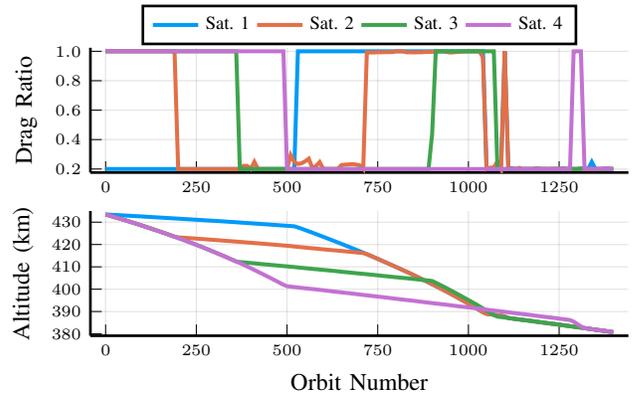}
    \caption{Scenario 1. Top: The control trajectories for the four satellites. Bottom: The altitude variation of the four satellites during the maneuver. Unlike in \cref{fig:lp_combined}, the control trajectories are not piecewise constant due to the on-line correction of disturbances.}
    \label{fig:line_combined_1}
\end{figure}

\begin{figure}[htb]
    \centering
    \includegraphics[width=\columnwidth]{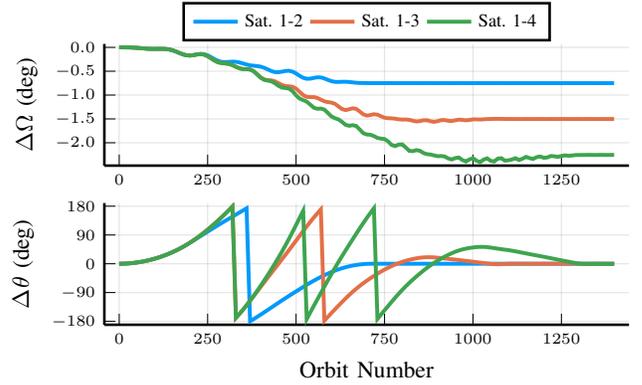}
    \caption{Scenario 1. Top: Right ascension of the ascending node difference with respect to the chief satellite. Bottom: Argument of latitude difference with respect to the chief satellite. All the satellite pairs reach the same final argument of latitude.}
    \label{fig:line_combined_2}
\end{figure}

The results of the first scenario are presented in \cref{fig:line_combined_1,fig:line_combined_2} and \cref{tab:results_line}.
The final orbit has a $385.5 \mathrm{km}$ altitude, eccentricity of $0.002$, and inclination of $51.49$ deg.
The top plot of \cref{fig:line_combined_1} shows the control trajectories for the four satellites.
The fourth satellite, the satellite that aims to reach the largest $\Delta \bar{\Omega}$, drives the overall differential drag required for the formation. 
The bottom plot shows the altitude variation for the four satellites; when the altitude rate is steeper, the satellite is in a high drag configuration. 
Contrarily, where the altitude rate is shallower, the satellite is in a low drag configuration.
\Cref{fig:line_combined_2} shows the AoL and RAAN difference for the three satellite pairs.
The difference is calculated with respect to the chief satellite.

\Cref{tab:results_line} reports the overall maneuver time, the final difference in the AoL and the RAAN, and the spherical distance between the chief satellite and the other satellite.
It takes three months to reach the final configuration, and the maximum distance between two satellites is 268.2 km. 

\begin{table}[htb]
\centering
\begin{tabular}{|l|c|l|l|l|}
\hline
\begin{tabular}[c]{@{}l@{}}Pair\end{tabular} &
  \multicolumn{1}{l|}{$t_f$,  months} &
  $\Delta \theta_f$, deg &
  $\Delta \Omega_f$, deg &
  \begin{tabular}[c]{@{}l@{}}Spherical\\ Distance, km\end{tabular} \\ \hline
Sat. 1 - 2 & \multirow{3}{*}{3}    &  0       & -0.75  &  89.26 \\ \cline{1-1} \cline{3-5} 
Sat. 1 - 3 &                       &  -0.008  & -1.5    & 178.7 \\ \cline{1-1} \cline{3-5} 
Sat. 1 - 4 &                       &   -0.213 & -2.25   & 268.2 \\ \hline
\end{tabular}
\caption{Results for Scenario 1}
\label{tab:results_line}
\end{table}

\subsubsection{Scenario 2 --- Square Formation} \label{sec:mpc_square}
The second scenario assumes that four satellites are deployed from an approximately sun-synchronous SpaceX Transporter launch, corresponding to an altitude of 550 km, an eccentricity of 0.005, and an inclination of $98^\circ$.
The goal is to maneuver the satellites to be distributed in AoL and RAAN to form the vertices of a square, as depicted in \cref{fig:scenario2}. 
For this scenario, the $\ell$ values are  0, 6, and 6, while the $\Delta \bar{\theta}_f$ are 0.03, 0, and 0.03.
The receding-horizon control policy is re-solved every five orbits over a time horizon of 5000 orbits.

\begin{table}[htb]
\centering
\begin{tabular}{|l|c|l|l|l|}
\hline
\begin{tabular}[c]{@{}l@{}}Pair\end{tabular} &
  \multicolumn{1}{l|}{$t_f$,  months} &
  $\Delta \theta_f$, deg &
  $\Delta \Omega_f$, deg &
  \begin{tabular}[c]{@{}l@{}}Spherical\\ Distance, km\end{tabular} \\ \hline
Sat. 1 - 2 & \multirow{3}{*}{10.8} &  10.78 & 0.005  &  1300.78 \\ \cline{1-1} \cline{3-5} 
Sat. 1 - 3 &                      &  -0.03 & 0.965  & 114.12 \\ \cline{1-1} \cline{3-5} 
Sat. 1 - 4 &                      &  10.85 & 0.95   & 1317.45 \\ \hline
\end{tabular}
\caption{Results for Scenario 2}
\label{tab:results_box}
\end{table}

\begin{figure}[htb]
    \centering
    \includegraphics[width=\columnwidth]{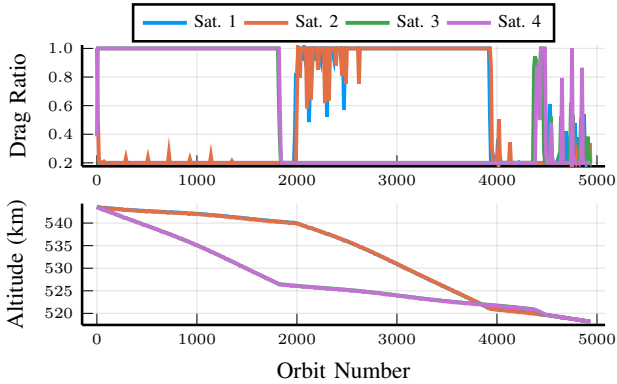}
    \caption{Scenario 2. Top: The control trajectories for the four satellites. Bottom: The altitude variation of the four satellites during the maneuver. Notice that unlike in \cref{fig:line_combined_1}, the satellites here change altitude in pairs, with only slight deviations to adjust for the desired argument of latitude difference.}
    \label{fig:box_combined_1}
\end{figure}

\begin{figure}[t]
    \centering
    \includegraphics[width=\columnwidth]{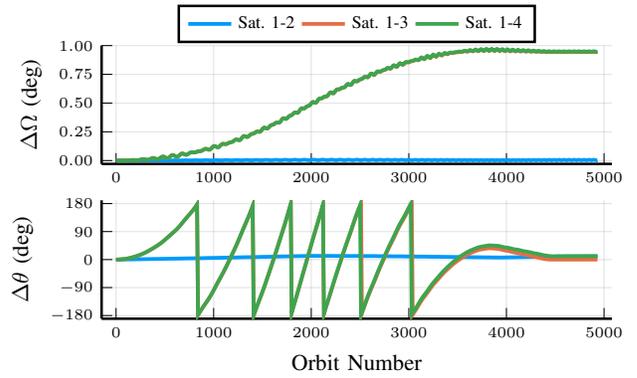}
      \caption{Scenario 2. Top: Right ascension of the ascending node difference with respect to the chief satellite. Bottom: Argument of latitude difference with respect to the chief satellite. Satellites 1-2 and satellites 3-4 reach the same right ascension of the ascending node, while satellites 1-4 and satellites 2-3 reach a comparable argument of latitude.}
    \label{fig:box_combined_2}
\end{figure}

The results of scenario 2 are presented in \cref{fig:box_combined_1}, \cref{fig:box_combined_2}, and \cref{tab:results_box}. 
The final orbit has a $518.1 \mathrm{km}$ altitude, eccentricity of 0.0022, and inclination of $98^\circ$. 
The top plot of \cref{fig:box_combined_1} shows the control input, and the bottom plot shows the altitude change for the four satellites. 
The plots in \cref{fig:box_combined_2} report the AoL and RAAN difference for the three pairs. 
As before, the difference is evaluated with respect to the chief satellite.
\Cref{tab:results_box} reports the overall maneuver time, the AoL and RAAN final differences, and the spherical distance between the chief satellite and every other satellite. 
This scenario takes longer than the first scenario; however, the results show that in less advantageous initial conditions a spacecraft formation with both along-track and cross-track separations can be established using our presented drag-based method. 
Furthermore, the algorithm is able to define the control trajectory in the presence of disturbances and 
modeling errors.  

The two scenarios have interesting differences from a mission-design viewpoint. 
Lower orbits, like the ISS orbit, result in faster natural orbital decay due to drag, reducing the possible altitude change.
However, the lower inclination for the ISS results in a larger $k_4$ and faster $\Delta \bar{\Omega}$ rate of $0.745^\circ$ per $2\pi$ revolution of$\Delta \bar{\theta}$.
Contrarily, deployment from the SpaceX Transporter allows a larger available overall altitude change but a smaller $\Delta \bar{\Omega}$ rate of 0.16 deg per $2\pi$ revolution of$\Delta \bar{\theta}$. 
This is why scenario 2 takes longer to complete than scenario 1.
The effectiveness of differential drag for cross-track formation control is highly dependent on orbital inclination.

\section{Conclusions} \label{sec:conclusion}
We have presented a novel control scheme that is able to maneuver a low-Earth orbit satellite formation in both along-track and cross-track directions without expending propellant.
We formulate the drag-based formation control problem as a linear program that can be solved very efficiently.
This allows it to be used in a receding-horizon manner, updating the control inputs and trajectory for a satellite once per orbit.
Our simulation results show that our method is robust to disturbances and unmodeled dynamics, and is viable for autonomous implementation on-orbit.
While we assume a known atmospheric density, in reality the atmospheric density in low-Earth orbit is widely varying.
An important extension of this work will be to accurately estimate the atmospheric drag; this estimate can be easily incorporated into our trajectory optimization formulation, and will ensure robust performance.
Our contributions can dramatically reduce the cost and complexity associated with deploying and managing multiple-plane satellite formations by eliminating the need for propulsion systems onboard the satellites.



\section{Acknowledgments}
This material is based upon work supported by the National Science Foundation under Grant No. 2111751.

A portion of this work was supported by the United States Department of Defense National Defense Science and Engineering Graduate Fellowship (NDSEG).

\balance
\bibliographystyle{IEEEtran}
\bibliography{refs.bib}

\end{document}